\documentclass[twocolumn]{revtex4}
\usepackage{amssymb}
\usepackage{amsmath}
\usepackage{graphicx}
\usepackage{epsfig}

\begin{document}

\title{\bf Superconducting energy gap in MgCNi$_3$ single crystals: Point-contact spectroscopy and specific-heat       measurements}

\author{Z. Pribulov\'{a},$^1$ J. Ka\v{c}mar\v{c}\'{i}k,$^1$ C. Marcenat,$^2$ P. Szab\'{o},$^1$ T. Klein,$^3$ A. Demuer,$^4$ P. Rodiere,$^3$ D. J. Jang,$^5$ H. S. Lee,$^5$ H. G. Lee,$^6$ S.-I. Lee,$^6$ and P. Samuely$^1$}

\affiliation{$^1$ Centre of  Very Low Temperature Physics at Institute of Experimental Physics, Slovak
Academy of Sciences, Watsonova 47, 04001 Ko\v sice, Slovakia\\
$^2$ CEA-Institut Nanosciences et Cryog\'enie/UJF-Grenoble 1, SPSMS, UMR-E 9001, LaTEQS, 17 rue des Martyrs, 38054 Grenoble, France\\
$^3$ Institut N\'eel, CNRS and Universit\'e Joseph Fourier, BP 166, 38042 Grenoble Cedex 9, France\\
$^4$ Grenoble High Magnetic Field Laboratory, CNRS, 38042 Grenoble, France\\
$^5$ Department of Physics, Pohang University of Science and Technology, Pohang 790-784, Republic of Korea\\
$^6$ Department of Physics, National Creative Research Initiative Center for Superconductivity, Sogang University,  Seoul 100-611, Republic of Korea}

\begin{abstract}
Specific heat has been measured down to 600 mK and up to 8 Tesla by the highly sensitive AC microcalorimetry on the MgCNi$_3$ single crystals with $T_c \approx$ 7 K. Exponential decay of the electronic specific heat at low temperatures proved that a superconducting energy gap is fully open on the whole Fermi surface, in  agreement with our previous magnetic penetration depth measurements on the same crystals. The specific-heat data analysis shows consistently  the strong coupling strength $2\Delta/k_BT_c \approx 4$. This scenario is supported by the direct gap measurements via the point-contact spectroscopy. Moreover, the spectroscopy measurements show a decrease in the critical temperature  at the sample surface accounting for the observed differences of the superfluid density deduced from the measurements by different techniques.
\end{abstract}
\maketitle
\section{Introduction}

Discovery of superconductivity in a cubic antiperovskite MgCNi$_3$ with the large ratio of Ni (60\% in molar ratio) at about 8 K \cite{He} was a surprise and it evoked a possible unconventional superconducting mechanism,     
where magnetic interactions may play an important role. Energy band calculations \cite{Mazin}  have shown that the density of states at the Fermi level is dominated by Ni $d$ states with a strong van Hove singularity yielding  a narrow and strong peak in the density of states just below the Fermi energy. This type of narrow energy
peak is typical for materials that display strong magnetic interactions. The peak was confirmed by x-ray spectroscopy
experiments \cite{Shein,Kim}, but its spectral weight was largely suppressed compared with the theoretical predictions. Attempts to introduce a long range magnetic order by  increasing  the DOS via doping  the Ni site has not been successful \cite{Rogacki}. Experimental indications of enhanced spin fluctuations in MgCN$_3$ have been found by the NMR investigations  by Singer et al. \cite{Singer} together with isotropic $s$-wave superconductivity.        
Several  papers have proposed a non conventional ({\it d}-wave) superconducting order parameter based on  experimental findings of a non-conventional critical current behavior \cite{Young} and the zero-bias tunneling conductance \cite {Mao}. 
The penetration depth distinctly exhibited a non-BCS low
temperature behavior \cite{Prozorov}. The previous reports on the specific heat show conventional $s$-wave superconductivity with phonon mediated pairing mechanism \cite{Mao,Lin,Shan,Walte,Shan2}, but at the same time an unusual low temperature behavior was observed as well. The latter effect was attributed either to the Schottky contribution  and/or the paramagnetism of unreacted impurities \cite{Mao,Lin} or to an electron-paramagnon interaction in MgCNi$_3$ itself \cite{Shan2,Walte}. W\"alte et al. \cite{Walte} have proposed a two-band/two-gap model to account for different sizes of superconducting gaps found by different techniques. The anomalous  point-contact Andreev-reflection spectra obtained by Shan et al. \cite{Shan3} were interpreted using a model in which point contact made on a BCS superconductor is in series with the Josephson junction due to polycrystalline character of the samples.  To resolve this controversial situation measurements on single crystals of good quality are highly desirable and they have recently appeared \cite{Leecrystals}. In our previous  studies of magnetic penetration depth on those crystals \cite{Diener} a fully open energy gap has been found in contrast to the results obtained on polycrystals. In our study, it was also found that the superfluid density extracted from the lower critical field was very different from that extracted from the tunnel diode oscillator measurements performed on the same sample. This discrepancy was related to the depletion of the critical temperature at the surface of the sample.

Here we present a detailed study of the high quality MgCNi$_3$ single crystals by specific heat ($C_p$) and point-contact spectroscopy (PCS) measurements. One of the aims of this work has been addressing the issue of differences which can appear between the bulk measurements ($C_p$) and surface measurements (PCS and previous penetration depth ($\lambda$) studies \cite{Diener}). That is why we used the same crystals or crystals  from the same batches as those measured for $\lambda$. Exponential decay of the electronic specific heat at low temperatures confirmed that a superconducting energy gap is fully open on the whole Fermi surface, in  agreement with our previous penetration depth measurements. The specific-heat data analysis shows consistently  the strong coupling strength $2\Delta/k_BT_c \approx 4$. This scenario is supported by the direct gap measurements via the point-contact spectroscopy. Moreover, the point-contact spectroscopy measurements show a decrease in the critical temperature  at the sample surface, accounting for the observed differences in the superfluid density deduced from the measurements by different techniques.

\section{Measurements}

Recently a long-standing problem of MgCNi$_3$ single-crystals preparation have been overcome. The tiny samples were fabricated in a high-pressure closed system. Details of the synthesis can be found elsewhere \cite{Leecrystals}. Using an X-ray micro analyzer it was proven that carbon deficiencies in stoichiometry were negligible. However, in contrast to polycrystalline MgCNi$_3$, which has usually local carbon deficiency, in these single crystals the Ni site was partly deficient. This was probably a reason for certain scattering in critical temperature among different crystals.  ${\it T}_cs$ of our crystals as measured by specific heat were found between 6 and 7.5 K. Single crystals  with a thickness of 0.1 mm have a rectangular shape and size of  about 0.25 x 0.15 mm$^2$.

Specific heat measurements have been performed  using an AC
technique, as described elsewhere \cite{Sullivan,Kacmarcik}. AC calorimetry technique consists of applying periodically modulated sinusoidal power and measuring the resulting sinusoidal temperature response.
In our case an optical fiber is used to guide the heating power emitted from the diode toward the sample. Absence of a contact heater reduces the total addendum to the total specific heat. The temperature of the sample is recorded by a thermocouple. A precise in situ calibration of the thermocouple in magnetic field was obtained from measurements on ultrapure silicon. The magnetoresistance of the Cernox thermometer was precisely inspected and corrections were included in the data treatment. Although an AC calorimetry is not capable to measure the absolute values of the heat capacity, it is a very sensitive technique for measurements of relative changes on minute samples and enable one to carry out
continuous measurements.  We performed measurements at
temperatures down to 0.6 K and in magnetic fields up to 8 T in the
$^3$He and $^4$He refrigerators.  

The point-contact spectroscopy measurements were performed in the $^4$He refrigerator. A standard lock-in technique was used to measure the differential resistance as a function of applied voltage on  point contacts. The micro-constrictions were prepared in situ by soft pressing of mechanically formed platinum tip to the surface of the sample. The special approaching system enabled both the lateral and vertical positioning of the tip by differential screw mechanism. Hence, our apparatus enables to change the place where the tip touches the sample. Moreover, by regulating the tip pressure it is possible to vary penetration of the tip into the sample surface.

\section{Results and discussion}

Figure 1 shows the temperature dependence of the total specific heat of the sample 
(plus addenda) in selected magnetic fields up to 8 T. \
The zero-field anomaly at the
transition is very sharp ($\Delta T_{c}\approx$ 0.15 K) indicating
the high quality and homogeneity of the single crystal sample, much improved in comparison to polycrystals. 
The positions of
the  specific-heat jump  gradually shift toward lower
temperatures for increasing magnetic field. The anomaly remains well resolved at
all fields, showing only little broadening at high fields.
Later we extended the measurements down to 0.6 K in a $^3$He
fridge, where the specific heat was measured at zero field and in 8 Tesla.  A field of 8 Tesla was
sufficient to completely suppress superconductivity down to about 4 K.
The normal-state specific heat has been obtained from fits to the 8-Tesla data between 4 and 12 K. 

\begin{figure}[t]
\begin{center}
\resizebox{0.45\textwidth}{!}{\includegraphics{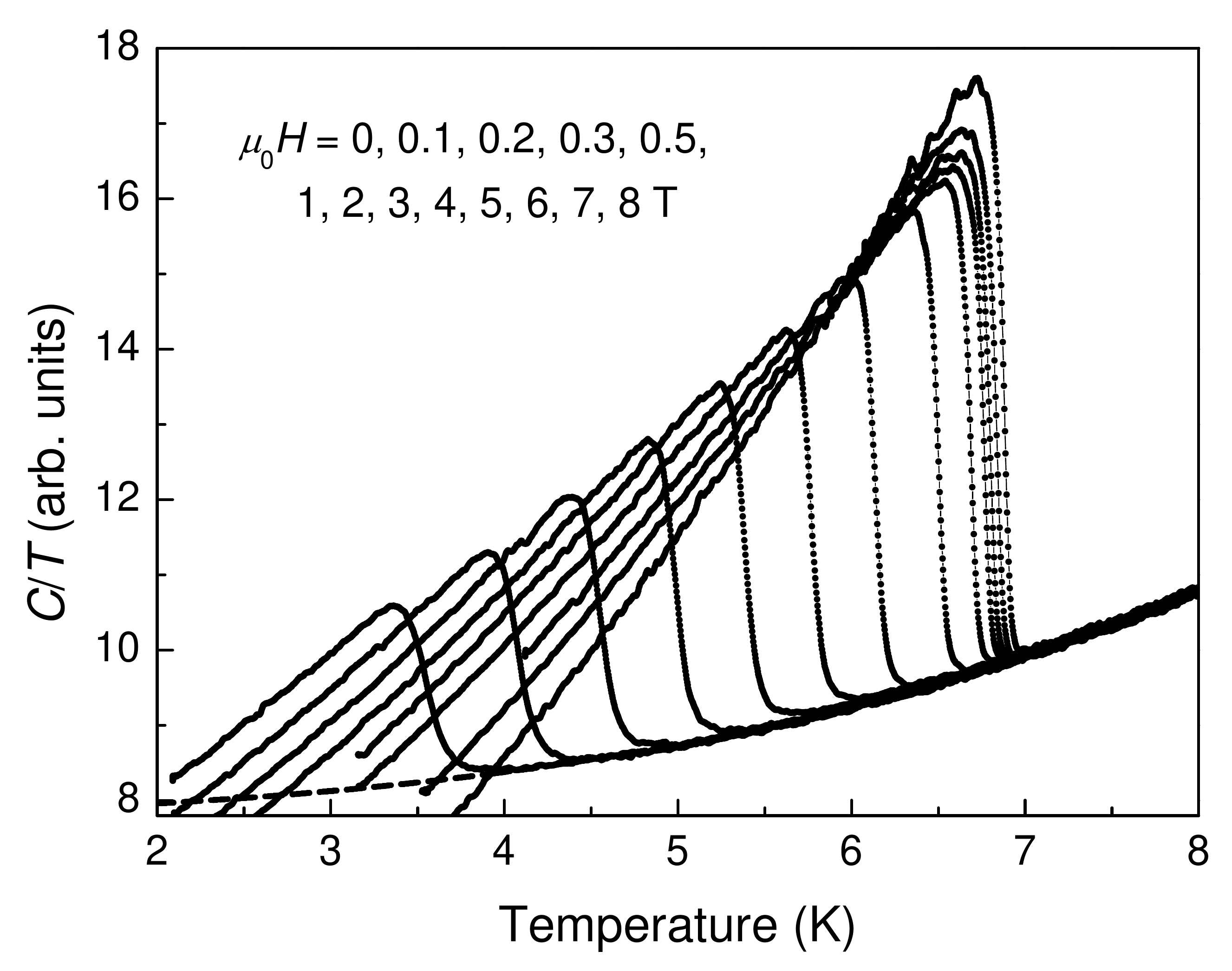}}
\caption{\label{label}Specific heat anomaly measured in magnetic fields. The zero-field measurements is the rightmost curve. Dashed line corresponds to the normal-state specific heat as calculated from the formula $C_n/T = a + bT^2 + cT^4$.}
\end{center}
\end{figure}

\begin{figure}[t]
\begin{center}
\resizebox{0.45\textwidth}{!}{\includegraphics{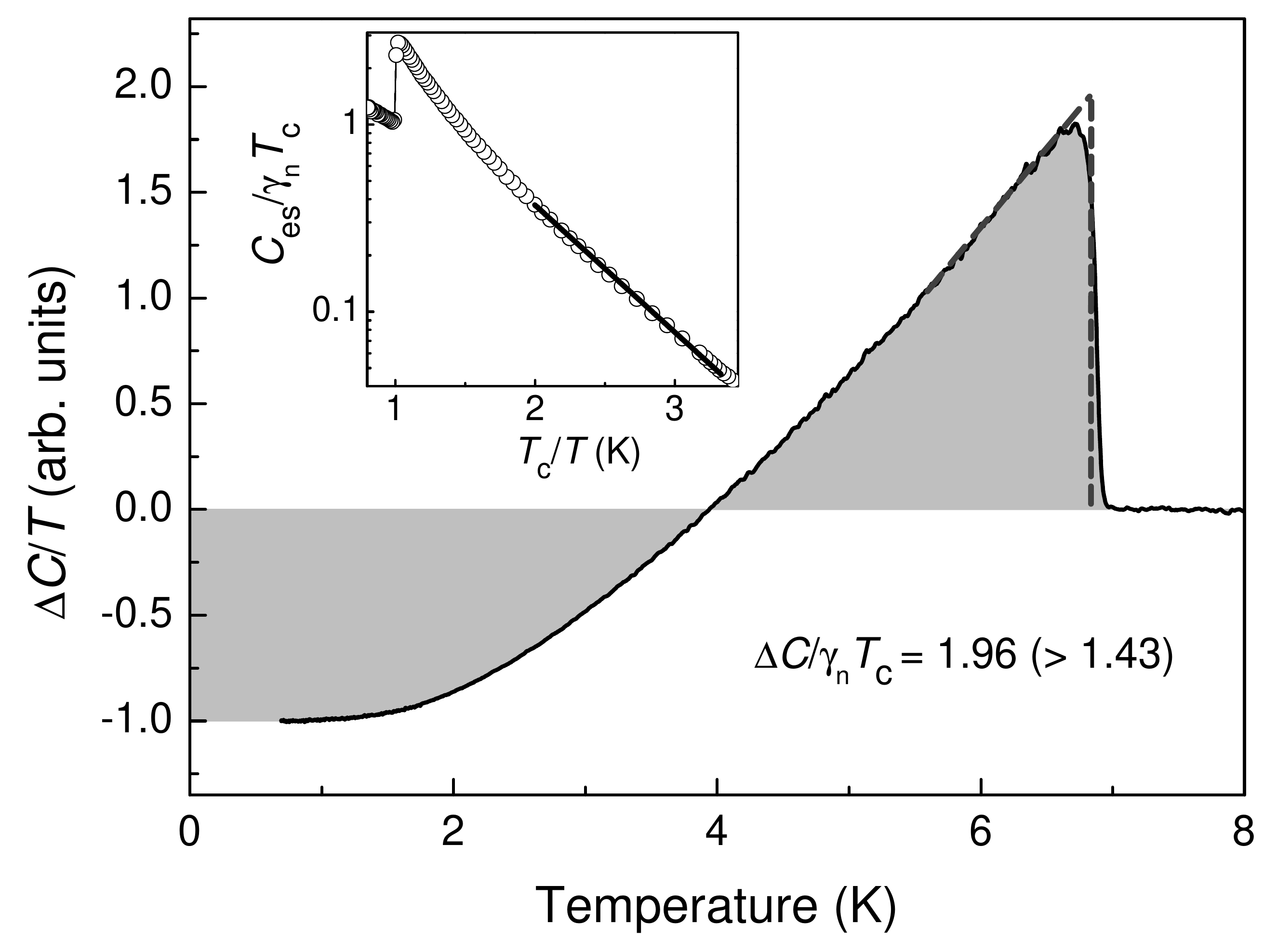}}
\caption{\label{label}Temperature dependence of specific heat in zero magnetic field. Dashed line is entropy conservation construction around critical temperature. Inset: exponential dependence of the electronic specific heat, full line represents the best fit of the exponential decay.}
\end{center}
\end{figure}

The normal-state specific heat of a non magnetic metal  consists of the electronic specific heat $\gamma_n T$ with the Sommerfeld coefficient $\gamma_n$ and the lattice part. At low temperatures the lattice part is usually described by the Debye model {\it C}$_{lattice}=\beta T^3$. However, in the case of MgCNi$_3$ the normal-state specific heat {\it C}$_n$ shows a systematic deviation from this description. In some cases, as in Ref.\cite{Mao,Lin},  a strong low temperature upturn of $C/T(T)$ is observed indicating a presence of Schottky anomaly probably due to magnetic impurities. In the case of polycrystalline samples of W\"alte et al. \cite{Walte} and Shan et al. \cite{Shan2} the deviations are much smaller and could be described either by higher phonon term ($\sim T^5$)
or by  additional electron-paramagnon interaction. In our case $C_n$ could also be fitted with $C_n=aT + bT^3 + cT^5$. Our $C_n$ comprises inevitably also the addenda but since all so far known specific heat measurements on MgCNi$_3$, independently on the form of the sample (polycrystals or single crystals) and the method of measurements, show the presence of  a low temperature upturn we believe it is intrinsic. We attribute it to the higher phonon term. This interpretation is supported by experimental observation of the softening of the lowest acoustic Ni phonon modes  below 200 K by Heid et al. \cite{Heid}. But our measurements cannot exclude a paramagnon contribution as well.

To derive the electronic specific heat we first subtracted the normal state specific heat {\it C}$_{n}$, i.e. we calculated $\Delta${\it C}({\it T})/{\it T} = {\it C}({\it T})/{\it T} - {\it C}$_{n}$({\it T})/{\it T}. By doing so we eliminate the addenda and phonon (eventually paramagnon) contribution. Figure 2 represents the resulting temperature dependence of $\Delta{\it C}/{\it T}$. The transition temperature obtained from entropy balance construction around anomaly (vertical dashed line in Fig.~2) is $T_c= 6.85$ K.  The entropy conservation required for a second order phase transition is fulfilled, as indicated by the hatched areas in Fig. 2 (see also the inset in Fig. 3). This check supports the  determination of the normal state specific heat and it verifies the thermodynamic consistency of the data. By the balance of entropy around the transition, the dimensionless specific-heat jump $\Delta C/\gamma_nT_c$ = 1.96 at $T_c$ is determined, where  $\gamma_n$=$\frac{C_n}{T}|_{0.6K} - \frac{C(H=0)}{T}|_{0.6K}$.  $\Delta C/\gamma_nT_c$ is an important measure of the electron coupling which is significantly stronger here than in the BCS weak coupling limit equal to 1.43. 

\begin{figure}[t]
\begin{center}
\resizebox{0.45\textwidth}{!}{\includegraphics{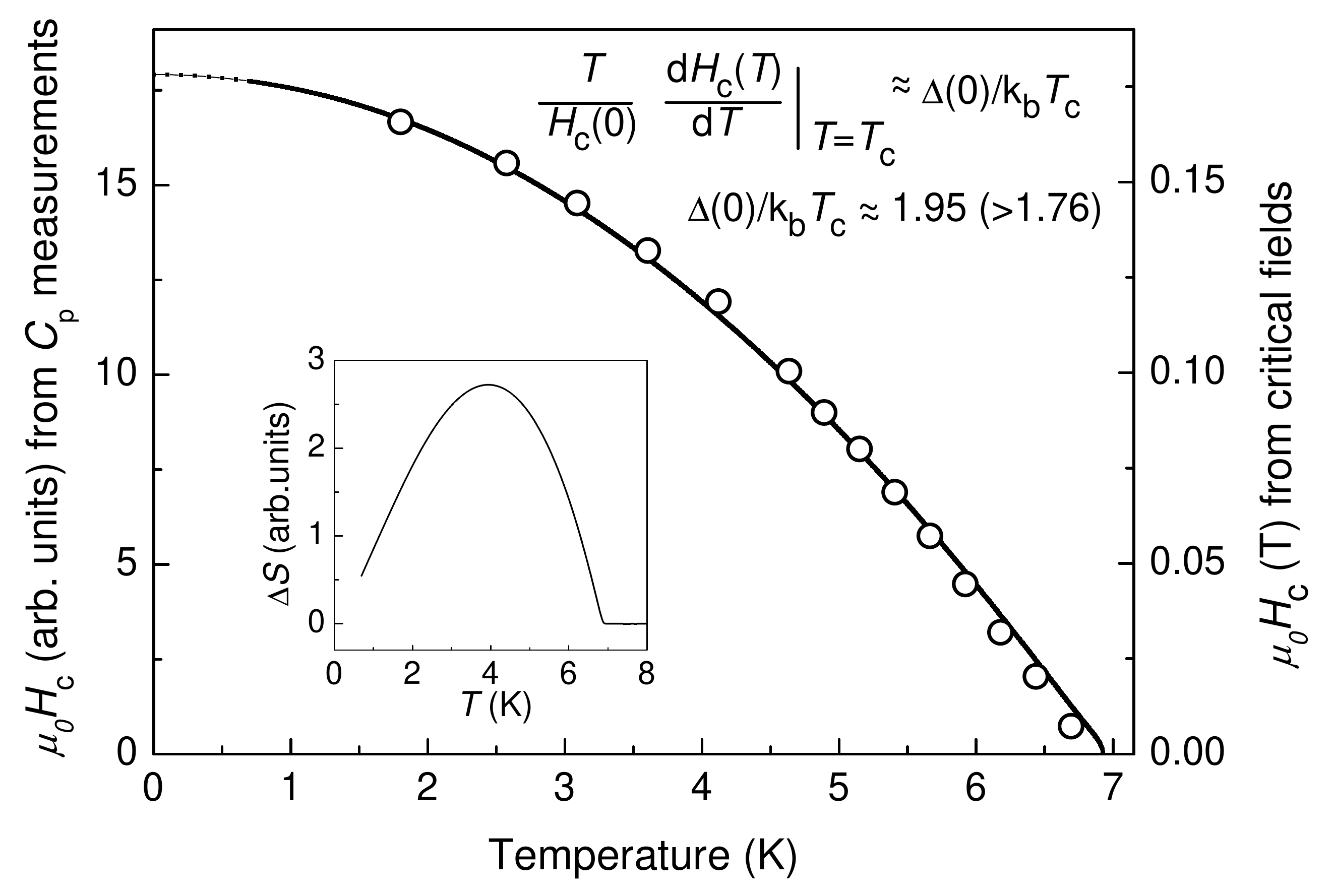}}
\caption{\label{label}Temperature dependence of thermodynamic critical field as derived from electronic specific heat (line), left axis applies. Right axis apllies to the $\mu_0H_c(T)$  as calculated from $H_{c1}$ found in Ref.\cite{Diener} and $H_{c2}$ from Fig. 4 (circles), see text for details. Inset: Difference of entropy between normal and superconducting state.}
\end{center}
\end{figure}

To further estimate the coupling strength we compared the electronic specific heat $C_{es}$/$T$ = $\Delta C/T$ + $\gamma_n$ of MgCNi$_3$ with the so-called alpha model \cite{Padamsee} based on the BCS theory. In this model the only adjustable parameter is the gap ratio  2$\Delta$/k$_{B}${\it T}$_{c}$. Our data could be well described by the model with a ratio 2$\Delta$/k$_{B}${\it T}$_{c}$ $\approx$ 4.2, which is much higher than the canonical value of 3.52 for the BCS superconductor. 

The inset of Fig. 2 displays the logarithm of  $C_{es}$ versus $T_c/T$. As shown, one obtains an exponential
dependence $C_{es} \propto c_1$exp(-$c_2$$T_c/T)$ for $T_c/T \geq 2$. This is a strong evidence that a full gap is present on the whole Fermi surface of MgCNi$_3$ and is in a full agreement with our previous penetration depth measurements \cite{Diener} on the same crystal. Exponential decay of the low temperature specific heat contradicts a presence of any non conventional order parameter in the system. The solid line  in the inset represents the best fit of the data  in the temperature range 2 - 3.5 K. The data could well be fitted by the expression 8.5exp(-1.57$T/T_c$) which shows the exponent higher than in the BCS prediction 8.5exp(-1.44$T/T_c$) valid  for this temperature range. This leads to the strong coupling ratio of $2\Delta$/k$_BT_c$ $\sim$ 3.84. The exponent 1.57 found in our experiment is even slightly higher than the value of 1.53 found by W\"alte et al. \cite{Walte}.

The thermodynamic critical field $H_c(T)$ contains also information about the coupling strength in the superconductor. $H_c$ can be determined from the electronic specific heat by double integration of the data. First of all, we calculate the difference of entropy between superconducting and normal state as $\Delta S(T`)$ = $\int_{T`}^{T_c}{(\Delta C/T}$)d$T$, i.e. the integral of the data from Fig.~2. Then, we get the thermodynamic critical field as $H_{c}^2$($T``$) = $8\pi$$\int_{T``}^{T_c}{\Delta S(T`)}$d$T`$, i.e. from the second integration. Figure 3 shows the resulting  temperature dependence of $H_c$ (line). The inset represents the difference of entropy between superconducting and normal state in MgCNi$_3$ calculated as explained above. Since the results of AC calorimetry measurements are in arbitrary units, such calculated $H_c$ is also in arbitrary units. Yet we can determine the ratio ($T/H_c$(0))(d$H_c$/d$T)|_{T\rightarrow T_c}$ which is close  to $\Delta(0)$/k$_BT_c$ \cite{Toxen}. Taking the value of $\mu_0H_c$(0) = 17.9 and derivative of $H_c$ in the vicinity of $T_c$ equal to 5.1, we get the coupling ratio $2\Delta$/k$_BT_c$ = 3.9, well in agreement with our previous estimates. 

To prove the consistency of our data we compared $H_c$ calculated from the  specific heat with  those calculated from the  lower and upper critical fields ($H_{c1}$ and $H_{c2}$, resp.) as  $\mu_0H_c$ = $\mu_0\sqrt{H_{c1}H_{c2}/(ln\kappa+0.5)}$, where the Ginzburg-Landau parameter $\kappa $ was determined from $H_{c1}$ and $H_{c2}$ as described in Ref.\cite{Brandt}. The temperature dependence of $H_{c1}$ was  taken from our previous work \cite{Diener} and the values of $H_{c2}$ are from Fig. 4 of this work. The resulting $\mu_0H_c(T)$, now in absolute units, is displayed in Fig. 3 by circles for which the right axis applies. The  data are in excellent agreement with those calculated by the double integration of the electronic specific heat. With the zero temperature coherence length $\xi(0) = 5.24$ nm determined from the upper critical field ($\xi = \sqrt{\Phi_0/2\pi\mu_0 H_{c2}}$ with the flux quantum $\Phi_0$) and $\lambda(0)= 230$ nm \cite{Diener} we get the Ginzburg-Landau parameter $\kappa(0)= 44$. All obtained critical parameters ($T_c, H_{c2}(0), H_{c1}(0), H_{c}(0)$) as well as the zero temperature coherence and penetration lengths are in a very good agreement with those determined by W\"alte et al. \cite{Walte} from the specific heat measurements on polycrystalline samples.

We have done the equivalent specific heat measurements and the data analysis on several other crystals with $T_c$ = 6 and 7.5 K. We found very similar results concerning the height of the specific heat jump $\Delta C(T_c)/\gamma_nT_c$ which was close to 2, as well as the value of coupling strength 2$\Delta$/k$_{B}${\it T}$_{c}$ to be close to 4.

\begin{figure}[t]
\begin{center}
\resizebox{0.45\textwidth}{!}{\includegraphics{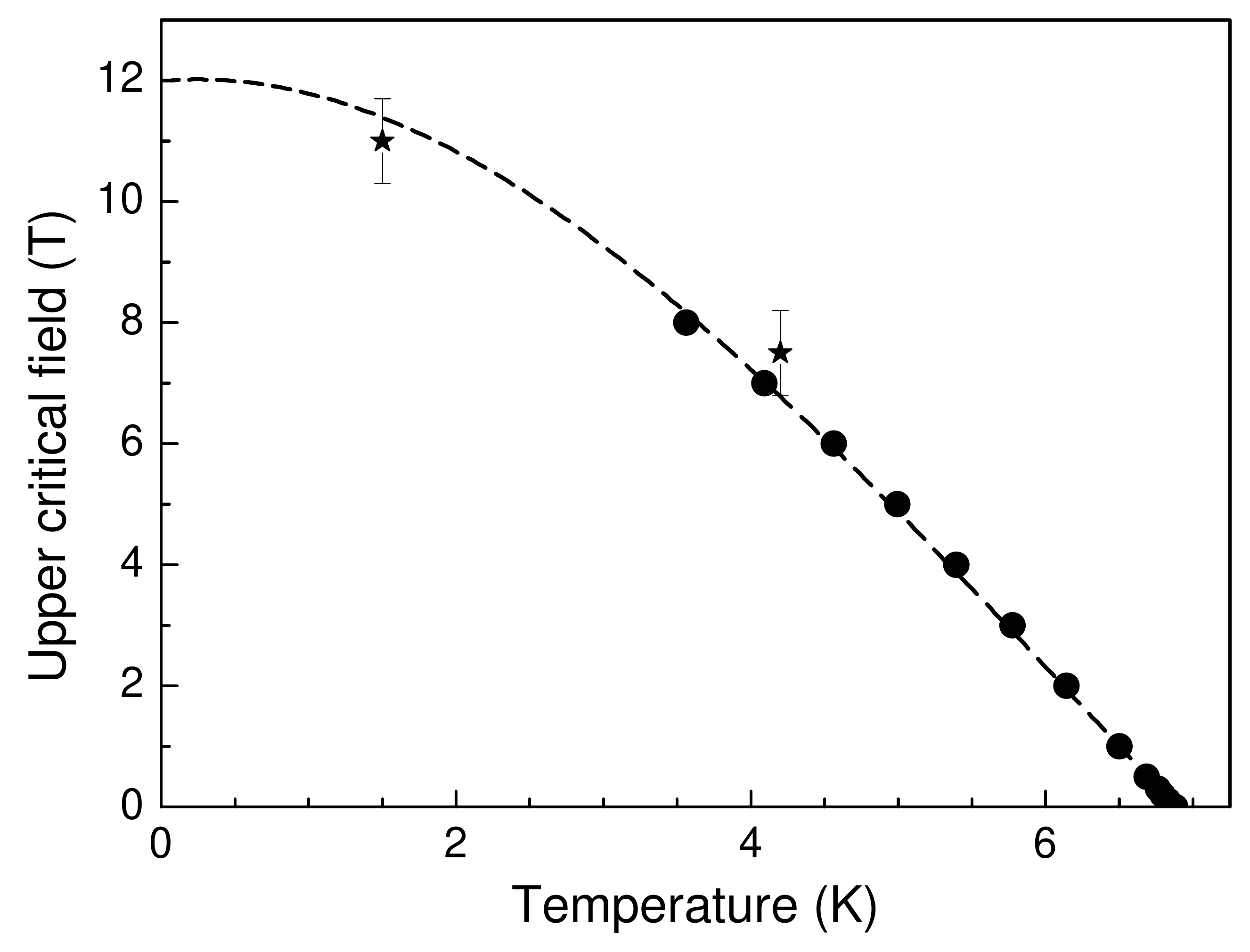}}
\caption{\label{label}Temperature dependence of the upper critical field in MgCNi$_3$ from specific heat measurements (circles) and point contact spectroscopy (stars). Dashed line is corresponding temperature dependence of {\it H}$_{c2}$ from WHH theory.}
\end{center}
\end{figure}

Figure 4 shows the upper critical field in MgCNi$_3$ derived from the specific heat measurements in magnetic field presented in Fig. 1. The mid-point of the transitions has been taken as a criterion to determine $H_{c2}$ for each magnetic-field measurement. Temperature dependence of $H_{c2}$ reveals a linear increase close to the critical temperature and gradual deviation from linearity at lower temperatures  as measured down to 3.5 K. Though measured only down to $T_c/2$ $H_{c2}(T)$  can be satisfactory described in the framework of the Werthamer-Helfand-Hohenberg (WHH) theory \cite{whh} as indicated by the dashed line in Fig. 4. WHH predicts that $H_{c2}(0)=0.693T_c\left(dH_{c2}/dT\right)_{T_{c}}$. 
With the slope $\mu_0\left(dH_{c2}/dT\right)_{T_{c}}=2.5$ T/K, we get $\mu_0H_{c2}(0)=12 $ T. Two points in Fig. 4 obtained from the point-contact spectroscopy measurements (see below) support this agreement even better. The Clogston paramagnetic critical field is the upper limit of the superconductivity at low temperature  and is given as $H_p(0)=\frac{1}{\sqrt{2}\mu_B}\Delta$ \cite{Clogston}, where $\mu_B$ is the Bohr magneton.  For our strong coupling case (with 2$\Delta$/k$_{B}${\it T}$_{c}=3.9$)  $\mu_0H_p(0)=14$ T is quite close the the observed values. 

One can calculate the Fermi velocity as $v_F=\pi\Delta\xi/\hbar = 3$x$10^4$ m/s (with our $\Delta = 1.15$ meV as determined above). This can be compared with the band structure calculations  of the Fermi velocities in the hole and electron subsystems in MgCNi$_3$ \cite{Walte}. The calculated hole Fermi velocity is $v_{F,h} = 12$x$10^4$ m/s and the electron one $v_{F,e} = 50$x$10^4$ m/s. For $H_{c2}(0)$ the slower Fermi velocity will play a dominant role and this is the hole one.  But in any way our calculated value is at least  four times smaller. This difference can be attributed to the strong electron-phonon renormalization factor $(1+\lambda_{e-ph})$. As introduced by McMillan and Werthamer \cite{McMillan} and further elaborated by Shulga and Drechsler \cite{Shulga}  the strong-coupling corrections lead to the modified  upper critical field as $H_{c2}^M(T)=H_{c2}(T)(1+\lambda_{e-ph})^n$, where $n \geq 2$. A very strong electron-phonon coupling constant $\lambda_{e-ph} \geq 2$ is then needed to explain the differences in the above mentioned Fermi velocities. As indicated in Ref.\cite{Walte} such a strong coupling would require a sizable depairing contribution to explain the low $T_c$ in MgCNi$_3$. One possible mechanism can be an existence of important electron-paramagnon interaction in the system as suggested in Refs.\cite{Shan2,Walte}.

Note however that a significant increase of the $H_{c2}$ value might also be induced by scattering effects in the so-called dirty limit in which case $H_{c2}$ becomes proportional to $1/(\xi .l)$, where $l$ is the electronic mean free path (see discussion in \cite{Walte}). Typically, our single crystals have a residual resistivity $\rho \approx$ 23 $\mu\Omega$cm \cite{Lee}. By evaluating the unrenormalized mean free path  $l=\frac{<v_F>}{\epsilon_0\rho\omega_p^2}$ with $<v_F>$ the average Fermi velocity (2.1x10$^5$ m.s$^{-1}$ \cite{Walte}) and $\omega_p$, the plasma frequency (3.17 eV \cite{Walte}), one gets a mean free path of few nm, suggesting the sample may be in or close to the dirty limit. 

Finally, a very large difference between the Fermi velocity deduced from band structure calculations and the one deduced from the upper critical field has also been obtained recently in Fe(Se,Te) \cite{Klein} (reaching in this case a factor of $\sim 20$). In this later system, this difference has been attributed to strong correlation effects in the normal states (see also \cite{Tamai}). To the best of our knowledge the role of correlations has not been addressed so far in MgCNi$_3$.  Note that both MgCNi$_3$ and Fe(Se,Te) share the similarity of having Fermi surfaces composed of both electron and holes pockets (3D sheets in MgCNi$_3$ instead of quasi 2D ones in iron based superconductors) and both systems are subjected to strong spin fluctuations. However,  it is generally assumed that those fluctuations lead to opposite effects in each system, being at the origin of the pairing mechanism in the so-called $s\pm$ model in iron based systems but strongly reducing the electron-phonon coupling constant in MgCNi$_3$. We show here that both systems also have very strong critical field (being close to the Pauli limit in MgCNi$_3$ and even limited by this Pauli field on a large part of the $H-T$ diagram in Fe(Se,Te)) and the similarity between those two systems probably deserve further works.

\begin{figure}[t]
\begin{center}
\resizebox{0.45\textwidth}{!}{\includegraphics{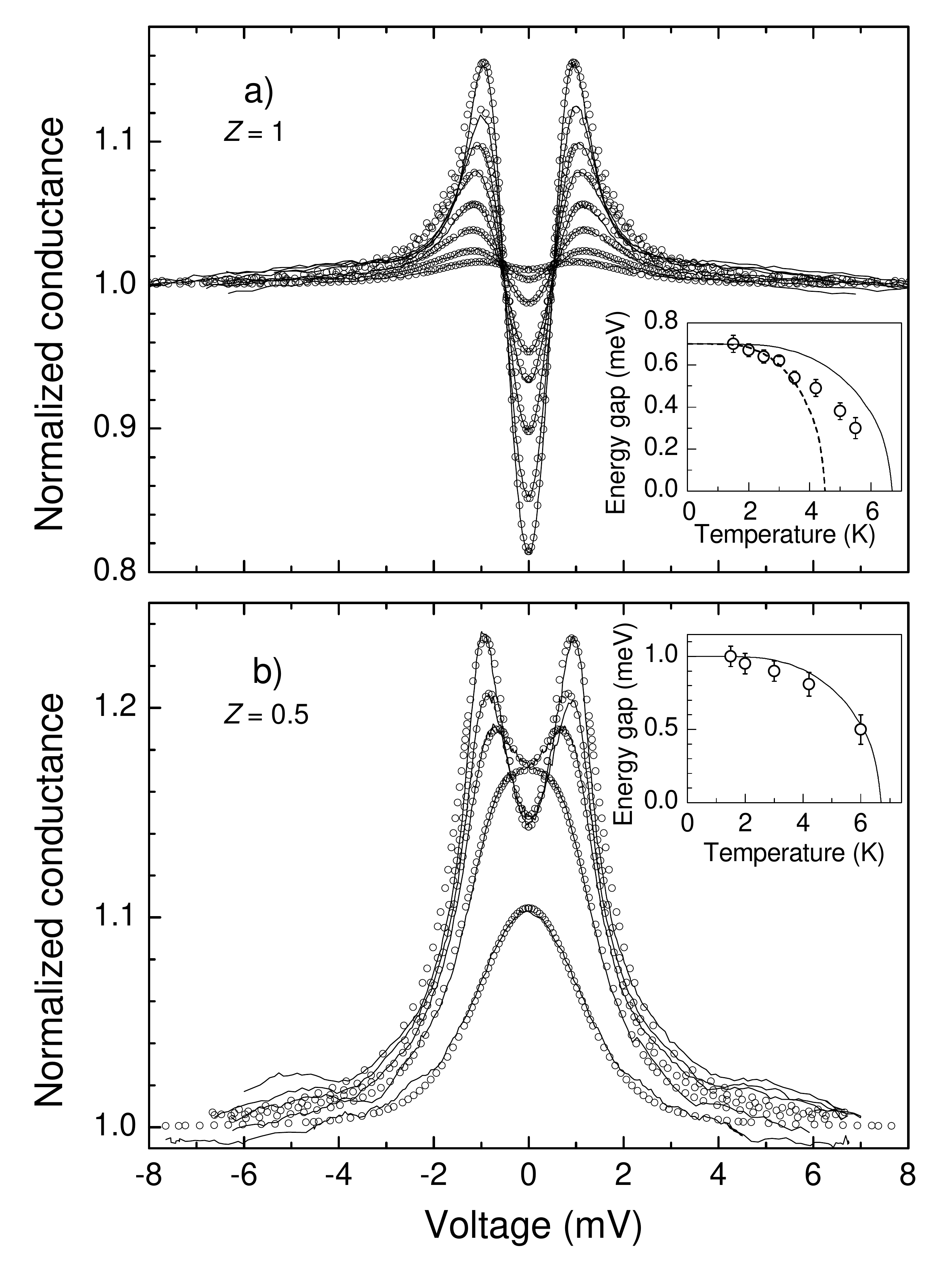}}
\caption{\label{label} a) Pt-MgCNi$_3$ point-contact spectra (solid lines) measured at 1.5 K, 2 K, 2.5 K, 3 K, 3.5 K, 4.2 K, 5 K, 5.5 K and fitting curves (open symbols).Inset: temperature dependence of energy gaps from fits to the spectra in the main figure (open symbols) and  BCS-like temperature dependences displayed by solid and dashed line for 2 different $T_c$s. b) Point-contact spectra of another junction (solid lines) measured at  1.5 K, 2 K, 3 K, 4.2 K, 6 K  with fitting curves (open symbols). Inset: temperature dependence of fitted energy gaps (open symbols) and  BCS-like temperature dependence.}
\end{center}
\end{figure}

The point-contact spectra have been measured mostly on a bigger crystal of  MgCNi$_3$ with the size of approximately 0.4x0.2x0.1 mm$^3$, but some  measurements were also done on the same crystal as used for the specific heat measurements with $T_c =6.85$ K. $T_c$ of the bigger crystal was 6.7 K, as determined locally by point-contact spectroscopy. The experimental differential conductance curves have revealed the typical characteristics of a single gap superconductor with a single pair of gap-like peaks. The measured point-contacts spectra were normalized to the conductance background found at higher energies above the superconducting  gap. This allowed for fits of the conductances to the point-contact model of Blonder, Tinkham and Klapwijk (BTK) accounting also for the spectral broadening  \cite{BTK} and to get the information about the energy gap $\Delta$, a parameter of the barrier strength $Z$ and a spectral broadening $\Gamma$. The barrier strength $Z$ was affected by adjustable pressure of the tip on the sample. Low-pressure junctions have  yielded more tunneling characteristics with the barrier parameters $Z \approx$ 0.6 - 1 while more Andreev-reflection characteristics with $Z < 0.6$ were found when the pressure was increased. The latter contacts revealed value of the gap scattered between 1.1 and 1.2 meV. Taking into account the $T_c \sim$ 6.7 K we obtain the coupling ratio of 2$\Delta$/k$T_c$ $\sim$ 3.8 - 4.2.  In some measurements on the junctions with more tunneling-like characteristics we have found the energy gap value at much smaller energies of around 0.7 meV.

\begin{figure}[t]
\begin{center}
\resizebox{0.43\textwidth}{!}{\includegraphics{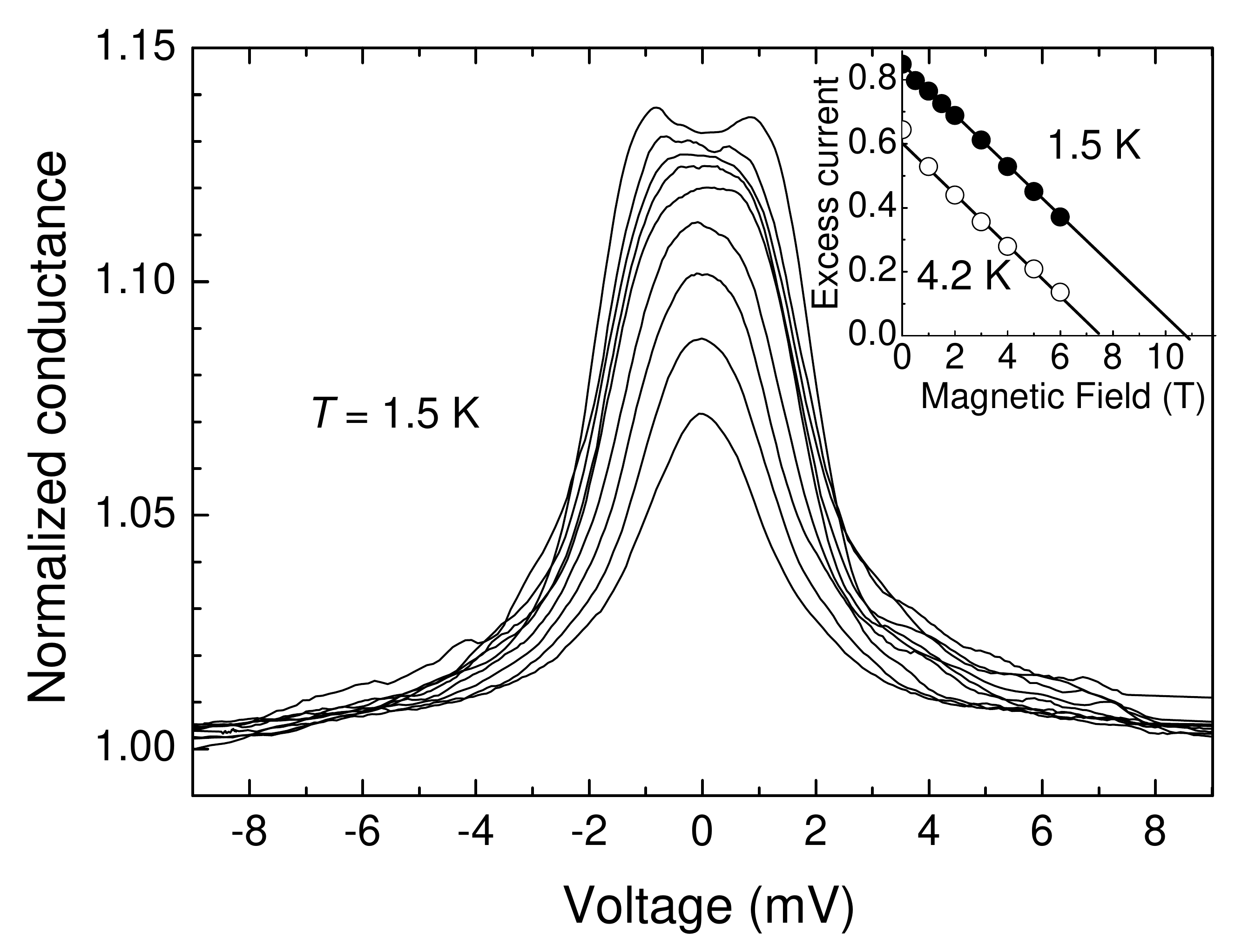}}
\caption{\label{label}Pt-MgCNi$_3$ point-contact spectra measured at 1.5 K in magnetic fields of 0 (topmost curve), 0.5, 1, 1.5, 2, 3, 4, 5 and 6 T. Inset: Determination of the upper critical field from magnetic field dependence of excess current.}
\end{center}
\end{figure}

In order to make more precise analysis some point contacts have been measured at different temperatures. Figure 5a displays the point-contact spectrum with a smaller gap (solid lines). The symbols represent the best fit to the BTK model at each particular temperature. Fitting parameters $\Gamma$ and $Z$ were first obtained for the spectrum at 1.5 K and then kept constant when fitting the curves measured at  higher temperatures. From the fit we found the superconducting gap $\Delta$ = 0.7 meV at 1.5 K with $\Gamma$ being on the order of 40$\%$ of $\Delta$ and $Z \sim$ 1. The inset shows  the temperature dependence of this  energy gap (open symbols). Surprisingly, the data displays a strong deviation from the BCS-type of temperature dependence with $\Delta (0)$ as a free parameter (solid line). Taking into account the  local critical temperature of the contact $T_c \sim$ 6.7 K one obtains the coupling  ratio  2$\Delta$/k$T_c$ $\sim$ 2.5. This value is much lower than the canonical BCS weak-coupling value of 3.52. Such a low value of the energy gap was also obtained on some point contacts measured by Shan et al. \cite{Shan3} on polycrystaline MgCNi$_3$. The fact that both a low energy gap and deviations from a typical BCS temperature dependence have been observed  on low pressure contacts, indicates that a degradated superconductivity on the surface of the sample has been at play. The degradated layer may have smaller $T_c$  than the bulk and the energy gap could reveal a tendency to close at lower temperatures than the bulk $T_c$, as is indicated by the dashed line in the inset,  yielding the  surface critical temperature of 4.5 K and the corresponding coupling constant of 2$\Delta$/k$T_c$ $\sim$ 3.5. The point contact probes the superconductivity on a distance of the coherence length from the junction. At increased temperatures the coherence length increases and adjacent layers deeper in the superconductung bulk with a higher $T_c$  will be probed. This causes that the measured gap is not following the dashed line, but it shows a finite value up to the bulk $T_c$ of 6.7 K. 
 
Figure 5b shows the spectra of another point contact (lines), now with a lower barrier strength, together with the fits at the selected temperatures (symbols). The fitting parameters  $\Delta$(1.5 K) = 1 meV, $\Gamma$ = 0.4 meV and $Z \sim$ 0.5 are found. The inset  displays the temperature dependence of this energy gap (symbols). Comparison with the BCS-like curve (line) shows that despite the fact that the ratio 2$\Delta$/k$T_c$ for this contact is close to 3.52,  there is still a deviation similar though smaller than in the previous case. This indicates that even the value of $\Delta(0)$ = 1 meV is an underestimate of the bulk energy gap \cite{remark} and that the value of 1.1 - 1.2 meV found on other junctions is closer the one related to the bulk phase with $T_c = 6.7$ K. Thus, the coupling ratio of 2$\Delta$/k$T_c$ $\sim$ 4.0 is suggested by our point-contact spectroscopy measurements, which is in a reasonable agreement with the value determined from the specific heat. 

In the previous measurements \cite{Diener} we  have observed a drastically different behavior between the superfluid
density at low temperatures extracted from the Hall probe and tunnel diode oscillator (TDO) measurements
performed on the same crystal of MgCNi$_3$. On the other hand, the difference has been vanishing near the common $T_c$. At low temperatures
 TDO measurements  probe only the sample surface while measurements by the Hall probe are sensitive to the bulk of the sample.  
Among possible explanations a systematic decrease in the critical temperature at the sample surface has been suggested. Lower $T_c$ at the surface would cause higher penetration depth that enters into the expression for superfluid density. We have estimated that 20$\%$  lower $T_c$ at the surface could explain the observed difference  in superfluid density.  The difference in $T_c$s indicated by the dashed and solid lines related to the small gap of $\Delta (0)$ = 0.7 meV (inset of Fig. 5a) is very close to this estimation supporting the explanation with a degradated superconductivity on the sample's surface. 

Figure 6 shows the normalized conductance spectrum of another point contact measured at 1.5 K at various magnetic fields. Apparently, the presented junction reveals quite significant contribution from the direct conductance, which is supported by the fit giving $Z\sim$ 0.3. The spectrum is very broadened with $\Gamma\sim$ 1.1 meV on the order of the value of the energy gap and the broadening is also responsible for a low intensity of the spectrum. Nevertheless, such a junction with a low barrier strength $Z$ can be used for determination of the excess current which can be approximated as the value of the area between the normalized conductance spectrum and unity: $I_{exc} \approx \int{(\frac{dI}{dV}-1)dV}$.  
The magnetic field dependences of $I_{exc}$ for 1.5  and 4.2 K of the junction are shown in the inset. In the both cases $I_{exc}$ decreases linearly with increasing magnetic field. A suppression of the excess current with increasing magnetic field is associated with the increasing number of vortices with cores representing a normal state area in the point contact. At the upper critical field $H_{c2}(T)$, at the normal state the excess current vanishes. By extrapolating $I_{exc} \rightarrow 0$,  the upper critical fields $\mu_0H_{c2}$ are found at 4.2 and 1.5 K. They are shown by stars  in Fig. 4. $\mu_0H_{c2}$(4.2 K) is in a good agreement with the specific heat determination and $\mu_0H_{c2}$(1.5 K) expands the experimental data to the lowest temperatures and proves the WHH type of temperature dependence. Importantly, the overall temperature dependence of the upper critical field is in a perfect quantitative agreement with the determinations from transport measurements done on the crystals from the same batch \cite{Lee}.

It is worth noticing that the magnetic field dependence of the excess current obtained on MgCNi$_3$ point contact behaves very differently from the case of MgB$_2$, a spectacular two-gap superconductor, where it decreases with two different subsequent slopes in 
accordance with different filling rates of the two gaps \cite{Samuely}. Then, the linear decrease of $I_{exc}(H)$ proves independently the presence of a single gap in the excitation spectrum of MgCNi$_3$.

\section{Conclusions}
Specific heat data obtained by AC calorimetry show sharp and well resolved superconducting transition in magnetic fields up to 8 Tesla. The results confirm very high quality of our 
MgCNi$_3$ single crystals used in the study. The low temperature electronic specific heat clearly reveals an exponential decrease - a strong evidence for $s$-wave superconductivity which is in a very good agreement with the previous penetration depth measurement on the same crystals. The ratio  2$\Delta$/k$_{B}${\it T}$_{c}$ $\approx$ 4 and high specific heat jump at the transition in zero field $\Delta${\it C}({\it T}$_{c}$)/$\gamma$$_{n}${\it T}$_{c}$ $\approx$ 1.96 confirmed the presence of strong coupling superconductivity in the system.
This scenario is supported by the direct gap measurements via the point-contact spectroscopy. Moreover, the spectroscopy measurements show a decrease in the critical temperature  at the sample surface accounting for the observed differences of the superfluid density deduced from the measurements by different techniques.

\acknowledgements

This  work  was supported  by the EC Framework Programme MTKD-CT-2005-030002, by the EU ERDF (European regional development fund) grant No. ITMS26220120005, by the Slovak Research and Development Agency,  under Grants No. VVCE-0058-07, No. APVV-0346-07, No. SK-FR-0024-09 and No. LPP-0101-06 and by the U.S. Steel Ko\v sice, s.r.o.  Centre of Low Temperature Physics is operated as the Centre of Excellence of the Slovak Academy of Sciences. We thank G. Karapetrov for careful reading of the manuscript.


\begin{thebibliography}{99}
\bibitem{remark} Remarkably, from the fits to the  low temperature $\lambda (T)$ measurements probing $\approx$ 100 nm below surface the energy gap around 1 meV has been determined \cite{Diener}.
\end{thebibliography}
\end{document}